%% file: 0_main.tex
\documentclass[conference]{IEEEtran}
\IEEEoverridecommandlockouts
\usepackage{cite}
\usepackage{amsmath,amssymb,amsfonts}
\usepackage{algorithmic}
\usepackage{graphicx}
\usepackage{textcomp}
\usepackage{xcolor}
\usepackage{url}
\def\BibTeX{{\rm B\kern-.05em{\sc i\kern-.025em b}\kern-.08em
    T\kern-.1667em\lower.7ex\hbox{E}\kern-.125emX}}
\input{preamble}

\usepackage{pbalance}
\begin{document}

\title{Evaluating False Alarm and Missing Attacks in CAN IDS
}

\author{\IEEEauthorblockN{ Nirab Hossain}
\IEEEauthorblockA{\textit{Department of Applied Mathematics} \\
\textit{University of Colorado Boulder}\\
Boulder, CO 80309, USA \\
nirab.hossain@colorado.edu}
\and
\IEEEauthorblockN{Pablo Moriano}
\IEEEauthorblockA{\textit{Computer Science and Mathematics Division} \\
\textit{Oak Ridge National Laboratory}\\
Oak Ridge, TN 37830, USA \\
moriano@ornl.gov}
}
\maketitle
\begin{abstract}
    \input{1_abstract}

\end{abstract}

\begin{IEEEkeywords}
CPS Security, Intrusion Detection, Controller Area Networks, Machine Learning Security, Adversarial Machine Learning.

\end{IEEEkeywords}

\input{2_intro}
\input{3_related_works}
\input{4_methods}

\input{5_results}
\input{6_conclusion}
\input{ack}

\bibliographystyle{IEEEtran}
\bibliography{sample}

\end{document}

%% file: preamble.tex
\usepackage{enumitem}
\usepackage{hyperref}

\usepackage{tikz}
\usepackage{fontawesome5}
\usetikzlibrary{arrows.meta,positioning,shapes.geometric,shapes.misc, shapes.symbols, calc, fit}
\tikzstyle{process} = [rectangle, fill=gray!7, minimum width=2cm, minimum height=1cm, text centered, draw=black]
\tikzstyle{decision} = [diamond, draw=black, text centered, minimum width=2cm, minimum height=1cm, aspect=2]
\tikzstyle{input} = [rectangle, rounded corners, draw=black, fill=red!10, minimum height=1cm, minimum width=2cm]
\tikzstyle{output} = [ellipse, draw=black, minimum height=1cm, minimum width=2cm]
\tikzstyle{startstop} = [rectangle, rounded corners, draw=black, dashed, minimum height=1cm, minimum width=2cm]
\tikzstyle{adversary} = [rectangle, draw=black, minimum height=1cm, minimum width=1.5cm, fill=orange!20]
\tikzstyle{arrow} = [thick, -{Latex[length=3mm, width=2mm]}]

\usepackage{multirow}
\usepackage{makecell}
\usepackage{booktabs}

\usepackage{graphicx} % Required for inserting images

\usepackage{xcolor}
\definecolor{dkgreen}{rgb}{0,0.6,0}
\definecolor{gray}{rgb}{0.5,0.5,0.5}
\definecolor{mauve}{rgb}{0.58,0,0.82}
\definecolor{Darkgreen}{rgb}{0.2,0.4,0.2}
\definecolor{Ltgreen}{rgb}{0,0.8,0}
\definecolor{Dred}{rgb}{0.7,0,0}
\definecolor{dred}{rgb}{0.5,0,0}
\definecolor{ORange}{HTML}{FF7F00}
\definecolor{mygreen}{RGB}{28,172,0}
\definecolor{mylilas}{RGB}{170,55,241}
\definecolor{listinggray}{gray}{0.9}
\definecolor{lbcolor}{rgb}{0.9,0.9,0.9}

\newcommand{\R}{\color{red}}

\newcommand{\dr}{\color{dred}}

\usepackage{float}
\usepackage{titlesec}
\titlespacing{\section}
{0pt}   % left indent
{6pt}   % space before
{4pt}   % space after

\titlespacing{\subsection}{0pt}{4pt}{2pt}

%% file: 1_abstract.tex
Modern vehicles rely on electronic control units (ECUs) interconnected through the Controller Area Network (CAN), making in-vehicle communication a critical security concern. Machine learning (ML)-based intrusion detection systems (IDS) are increasingly deployed to protect CAN traffic, yet their robustness against adversarial manipulation remains largely unexplored. We present a systematic adversarial evaluation of CAN IDS using the ROAD dataset, comparing four shallow learning models with a deep neural network–based detector. Using protocol-compliant, payload-level perturbations generated via FGSM, BIM and PGD, we evaluate adversarial effects on both benign and malicious CAN frames. While all models achieve strong baseline performance under benign conditions, adversarial perturbations reveal substantial vulnerabilities. Although shallow and deep models are robust to false-alarm induction, with the deep neural network (DNN) performing best on benign traffic, all architectures suffer significant increases in missed attacks. Notably, under gradient-based attacks, the shallow model extra trees (ET) demonstrates improved robustness to missed-attack induction compared to the other models. Our results demonstrate that adversarial manipulation can simultaneously trigger false alarms and evade detection, underscoring the need for adversarial robustness evaluation in safety-critical automotive IDS.

% Modern vehicles rely heavily on electronic control units (ECUs) connected through the Control Area Networks (CAN), making in-vehicle communication a key target for cyberattacks. machine learning based intrusion detection systems (IDS) offer an increasingly important defense layer, yet their resilience to adversarial manipulation remains insufficiently examined. This work evaluates the adversarial robustness of four shallow models and a sequential deep learning (DL)-based model trained on the ROAD dataset to detect fuzzing and fabrication attacks by generating FGSM, BIM, and PGD perturbations, manipulating both benign and malicious frames under realistic CAN domain constraints. While all models achieve high Matthews correlation coefficient (MCC) under benign conditions, both shallow and DL-based models demonstrate greater resilience in false positives (FPs), they still suffer substantial increases in false negatives (FNs) under structural perturbations; even DL-based models experience catastrophic degradation under gradient-based attacks, producing 100\% attack success rate (ASR). These findings highlight how adversarial manipulation can both trigger false alarms and cause IDS to miss attacks, underscoring the need for adversarial evaluation frameworks and more robust model designs for safety-critical automotive systems.

%% file: 2_intro.tex
\section{Introduction}
Modern vehicles rely on numerous electronic control units (ECUs) to coordinate functions ranging from powertrain and braking to steering and body control~\cite{douss2023state}. Communication among these components is predominantly supported by the Controller Area Network (CAN), which remains the most widely deployed in-vehicle communication protocol in production vehicles~\cite{wu8688625}. Designed decades ago for reliable message exchange in closed automotive environments%{\B [CAN design goals and origin]}
~\cite{douss2023state, wu8688625}, CAN has persisted across vehicle generations due to its simplicity, low cost, and strong backward compatibility~\cite{douss2023state, wu8688625}%[persistence and backward compatibility]
. Although higher-bandwidth alternatives such as automotive Ethernet are increasingly introduced~\cite{eth7393668}, %[emergence of alternative in-vehicle networks], 
CAN continues to underpin many safety-critical subsystems and is unlikely to be phased out in the near term~\cite{ li2020survey, wu8688625}. %[mixed architectures and continued deployment] 
As a result, understanding CAN’s properties and limitations remains essential for assessing the reliability and security of modern automotive systems.

% 2. CAN vulnerabilities and high impact exploits 
Despite its widespread adoption, CAN was not designed with security as a primary objective~\cite{douss2023state, li2020survey, wu8688625, jo9439954}.
%[CAN design assumptions]
It lacks fundamental protection mechanisms such as authentication, encryption, and source verification, operating under the assumption of a trusted in-vehicle environment~\cite{jo9439954}.
%[CAN security limitations]. 
As vehicles have become increasingly connected through wireless interfaces, diagnostic ports, and external services, this assumption no longer holds~\cite{checkoway2011comprehensive, koscher2010experimental}. 
%[increased vehicle connectivity]. 
Prior work has shown that adversaries with CAN-bus access can inject, modify, or replay messages, enabling attacks that affect both vehicle functionality and safety~\cite{checkoway2011comprehensive}. %[CAN attack feasibility].
These attacks can impact critical systems such as braking, steering, and powertrain control, with consequences ranging from denial of service to loss of driver control~\cite{miller2015remote, checkoway2011comprehensive}. %[high-impact automotive exploits]. 
Consequently, the combination of protocol-level trust assumptions and safety-critical traffic has made in-vehicle networks an attractive target for high-impact exploits.

% 3. IDS for CAN: signature and ML-based: shallow vs DL 
To mitigate these risks, intrusion detection systems (IDS) have been widely proposed for CAN-based networks~\cite{wu8688625}. %[IDS for CAN overview]. 
Early approaches relied on rule-based or signature-driven techniques that detect known attack patterns with low computational overhead~\cite{Mter2010ASA, sig9401087}, %[signature-based CAN IDS]
but they struggle to generalize to unseen attacks and require frequent updates. To address these limitations, machine learning (ML)–based IDS have gained prominence by leveraging statistical patterns and learned representations of normal CAN traffic~\cite{wu8688625}. %[ML-based CAN IDS].
Initial solutions employed shallow models such as decision trees, support vector machines, and ensemble methods\cite{canf}, %[shallow ML IDS], 
offering interpretability and modest training requirements. More recent work has adopted deep learning (DL) architectures, including convolutional and recurrent models, to capture temporal dependencies and complex correlations across CAN messages~\cite{shahriar2023canshield, marfo2024detecting}. 
%[deep learning CAN IDS; for example Shahriar et al. 2023 and Marfo et al. 2025]. 
These DL-based approaches have demonstrated strong detection performance in controlled settings, accelerating their adoption in automotive IDS research.

% 4. the issues of adversarial attacks on ML-based CAN-IDS 
Despite their empirical success, ML-based IDS introduce vulnerabilities that manifest in distinct and equally problematic failure modes~\cite{fi15020062, apruzzese2022adversarial}. %[adversarial ML on IDS]. 
One failure mode involves false alarms, where benign CAN traffic is misclassified as malicious due to subtle input perturbations~\cite{aloraini2024adversarial}. %[false positives in ML-based IDS]. 
In safety-critical systems, excessive false positives can degrade reliability, trigger unnecessary mitigation actions, and erode trust in the IDS. A complementary failure mode arises when adversarial manipulations cause false negatives, allowing malicious CAN messages to evade detection~\cite{longari2023evaluating}. %[evasion attacks on CAN IDS]. 
Unlike rule-based systems, learned decision boundaries can be intentionally influenced through carefully crafted, protocol-compliant inputs~\cite{goodfellow2015explaining}%[adversarial examples]
, enabling attackers to induce either outcome without violating CAN specifications. These dual failure modes expose fundamental robustness challenges for ML-based CAN IDS under realistic adversarial conditions.

% 5. High-level summary of the novelty of the work and contrasts with the state of the art. conclude with bullet points with the specific contributions.
Building on these observations, this work evaluates the adversarial robustness of CAN-bus IDS by both separately and jointly examining false alarms and missed attacks as coupled adversarial failure modes that determine IDS reliability under attack~\cite{fi15020062}. We focus on protocol-compliant, payload-level manipulations that preserve CAN frame validity while influencing IDS decisions at inference time~\cite{aloraini2024adversarial}. 
%[protocol-compliant adversarial CAN manipulation]. 
Unlike prior studies that consider false alarms or evasion in isolation~\cite{aloraini2024adversarial, longari2023evaluating}, we adopt a unified evaluation framework applied consistently across representative ML-based IDS families. By grounding the analysis in realistic CAN attack scenarios from the ROAD dataset~\cite{Verma2024RoadDataset}, our study enables a more faithful assessment of adversarial effects under practical traffic and attacker conditions. This paper makes the following contributions:
\begin{itemize}[leftmargin=*]
\item We present a unified adversarial evaluation of CAN-bus IDS that both separately and jointly considers false alarms (false positives) and missed attacks (false negatives) as coupled failure modes.
\item We design and apply protocol-compliant, payload-level adversarial perturbations and evaluate them against realistic CAN attack scenarios from the ROAD dataset.
\item We conduct a comparative analysis across representative ML-based IDS families, including shallow and DL models, to characterize how architectural choices influence adversarial robustness.
\item We illustrate adversarial failure modes that are not captured when IDS performance is assessed only on clean test data.
\item To support reproducibility and further research, we make our implementation, experimental pipeline, and evaluation code publicly available at~\cite{gitrepo}.
\end{itemize}

%% file: 3_related_works.tex
\section{Related Works} \label{sec:related work}

IDS on the automotive CAN bus has been widely studied, with increasing emphasis on ML-based IDS. Early approaches relied on statistical or rule-based techniques to detect deviations from expected CAN traffic patterns~\cite{osti_1765478,Mter2010ASA, sig9401087}. While computationally efficient, these methods are typically limited to known attack signatures and struggle to generalize to unseen behaviors. Consequently, recent research has shifted toward data-driven IDS models. Sun \textit{et al.}~\cite{9521818}, for example, demonstrated that hybrid deep architectures combining CNNs and LSTMs with attention mechanisms can capture temporal dependencies in CAN traffic for anomaly detection. Beyond this, a broad range of DL architectures, including CNNs and GRUs, as well as ensemble methods, have been explored. In parallel, unsupervised techniques such as clustering and PCA, together with classical classifiers including SVMs, k-NN, and decision trees, have been proposed for CAN anomaly detection~\cite{canf,li2020survey}. Wu \textit{et al.}~\cite{wu8688625} provide a comprehensive survey of in-vehicle IDS approaches, highlighting the diversity of modeling choices and evaluation practices in the field.

At the same time, adversarial machine learning (AML) has emerged as a critical concern for security-sensitive ML systems. Seminal work by Goodfellow \textit{et al.}~\cite{goodfellow2015explaining} and Carlini and Wagner~\cite{carlini2017towards} established that carefully crafted gradient-based perturbations can cause neural networks to misclassify inputs with high confidence. In the broader cybersecurity domain, He \textit{et al.}~\cite{apruzzese2022adversarial} reviewed adversarial attacks on ML-based IDS and showed their effectiveness across a wide range of settings. However, within the automotive domain, adversarial research has largely focused on perception and autonomous driving tasks, with comparatively limited attention given to the robustness of in-vehicle IDS~\cite{mbow2021evaluating}.

A smaller body of work has begun to address adversarial attacks targeting automotive IDS directly. Pacheco \textit{et al.}~\cite{pacheco2021adversarial} applied FGSM, JSMA, and CW attacks to IDS, primarily analyzing adversarially induced false negatives. Zenden \textit{et al.}~\cite{zenden2023resilience} evaluated FGSM-based attacks on CAN IDS, examining transferability, adversarial training, and standard performance metrics such as FPR and FNR, but did not consider stronger multi-step attacks or provide a detailed analysis of failure modes under protocol-constrained perturbations. Longari \textit{et al.}~\cite{longari2023evaluating} proposed an online, oracle-guided evasion strategy that perturbs CAN payload bits in attacks such as Drop, Fuzzy, Replay, and Seamless Change, studying IDS robustness under different attacker knowledge assumptions. More recently, Aloraini \textit{et al.}~\cite{aloraini2024adversarial} investigated adversarial attacks on ML-based CAN IDS by applying FGSM, BIM, PGD, and decision-tree–based attacks to benign traffic, showing that adversarial perturbations can reliably induce false alarms across multiple models.

In \emph{contrast} to prior work, our study evaluates both shallow and DL-based CAN IDS using a unified adversarial framework grounded in the comprehensive ROAD dataset. By applying protocol-compliant adversarial attacks, including FGSM, BIM, and PGD, we provide a more architecture-sensitive and realistic assessment of IDS robustness. Crucially, we examine adversarially induced false alarms and missed attacks both in isolation and in combination, enabling a more complete characterization of IDS reliability in safety-critical in-vehicle network settings.

Dataset choice is central to IDS evaluation. While several CAN intrusion datasets exist, many are limited in duration or attack diversity. For example, the Car-Hacking dataset %~\cite{kang2016intrusion} 
contains only a small number of attack types and short driving traces. In contrast, the ROAD dataset~\cite{Verma2024RoadDataset} offers over 3.5 hours of real vehicle CAN traffic, comprising 12 benign captures and 33 attack captures, including random fuzzing floods, targeted fabrication attacks, and ECU masquerade scenarios. We therefore adopt ROAD to ensure that our evaluation reflects diverse driving conditions and adversarial behaviors. %encountered in practice.

%% file: 4_methods.tex
%\section{Materials and Methods}
\section{Materials and Methods}

\subsection{CAN}

CAN is the primary in-vehicle communication bus used to interconnect multiple ECUs~\cite{Moriano2025OnlineCANIDS}. It employs a broadcast, message-based protocol in which each frame, illustrated in Fig.~\ref{fig:can_frame}, consists of an identifier (ID), a data length code (DLC), up to eight data bytes, and error-checking fields. Medium access is governed by priority-based arbitration, where frames with lower numerical IDs are granted access to the bus before higher-priority messages.
\input{Figures/figure_CAN}

In classical CAN, the \emph{payload} corresponds to the raw data field comprising up to eight bytes (D0–D7), represented as hexadecimal values in CAN logs. %In this work, these bytes are not interpreted semantically. Instead, they
Those data bytes are treated directly as numerical features derived from observed CAN traffic. Under this representation, each frame contributes its eight data bytes as independent features, enabling both benign and malicious behaviors to be analyzed at the level of raw payload values.

Within our methodology, adversarial manipulation is restricted to the payload bytes.
%, while the CAN ID and timing information are preserved. 
This design choice reflects an attack surface in which an adversary alters message contents without violating protocol structure or disrupting normal bus arbitration. Although CAN provides efficient and reliable real-time communication, it lacks fundamental security mechanisms such as authentication and encryption. Consequently, a compromised ECU can inject, modify, or replay messages that appear legitimate to other nodes on the network. These limitations have motivated extensive research on intrusion detection systems for CAN networks, including both shallow and DL–based unsupervised approaches~\cite{Moriano2025OnlineCANIDS, osti_1765478, wu8688625, jo9439954}.

% {% 3.2
% \subsection{IDS Architecture}
% }
% Our IDS operates at the frame level, using supervised machine learning models to classify each CAN message independently based on its CAN ID, DLC, and payload byte features, without relying on sliding windows or temporal aggregation (prepared as described in \ref{sub:prep}). We employ five classifiers, i.e., DT, RF, ET, XGB and DNN. DT provides simple hierarchical rules but is prone to overfitting or underfitting depending on depth, while RF and ET mitigate variance through ensembling and typically yield stronger and more stable baselines. XGB further enhances performance via gradient-boosted trees that capture {\B complex feature interactions across payload bytes}. The DNN {\B consists of four fully connected hidden layers with 16 neurons each (ReLU activation), followed by a single sigmoid output neuron for binary classification.} Including these models enables a direct comparison between shallow and DL methods under benign and adversarial evaluation {\B settings}.\\

\subsection{IDS Architecture}

Our IDS operates at the frame level and uses supervised ML models to classify each CAN message independently based on its payload byte features. The classification is performed without relying on sliding windows or temporal aggregation, with feature preparation conducted as described in~\ref{sub:prep}. This design isolates per-frame decision behavior and enables a controlled analysis of model responses under benign and adversarial conditions.

We consider five classifiers: decision tree (DT), random forest (RF), extra trees (ET), XGBoost (XGB), and a deep neural network (DNN). DT provides simple hierarchical decision rules but is sensitive to model depth, which can lead to overfitting or underfitting. RF and ET address this limitation by aggregating multiple randomized trees, reducing variance and yielding stronger and more stable baselines. XGB further improves performance through gradient-boosted trees that capture complex feature interactions across payload bytes. The DNN consists of four fully connected hidden layers with 16 neurons each using ReLU activations, followed by a single sigmoid output neuron for binary classification. Following the work by Aloraini \textit{et al.}~\cite{aloraini2024adversarial}, these models enable a direct comparison between shallow and DL–based IDS approaches under both benign and adversarial evaluation settings.

% \subsection{Threat Model}
% {\B We assume an external attacker with access to the CAN bus who can inject fabricated messages while imitating legitimate ECUs (gray-bus scenario). For adversarial evaluation, we assume the attacker has white-box knowledge of the IDS, enabling the computation of gradient-based perturbations.} The adversary's aim is to avoid detection by the IDS while causing unauthorized effects (spoofing sensor readings or disabling alerts). {\B The adversary may also exploit false positives by triggering benign frames to be flagged as malicious, disrupting normal operations and eroding trust in the IDS over time.} 
% \input{Paper/Tables/table_stats}

\subsection{Threat Model}

We assume an external adversary with access to the CAN bus who is capable of injecting fabricated messages while imitating legitimate ECUs, corresponding to a gray-bus attack scenario. For adversarial evaluation, we further assume that the attacker has white-box knowledge of the IDS, which enables the computation of gradient-based adversarial perturbations. The adversary’s primary objective is to exploit false positives by manipulating benign frames so that they are flagged as malicious, thereby disrupting normal vehicle operation and gradually eroding trust in the IDS as well as to
evade detection by the IDS while inducing unauthorized effects, such as spoofing sensor readings or suppressing alerts. This threat model captures both false-alarms and evasion–oriented adversarial objectives relevant to safety-critical CAN environments.

% %3.4
% \subsection{Dataset}
% We have used the ROAD CAN IDS dataset \cite{Verma2024RoadDataset} which is a recent, high-fidelity collection of CAN bus data containing both normal driving sessions and a diverse set of attack scenarios. It comprises over 3.5 hours of CAN traffic from a real vehicle under dynamometer testing conditions, including 12 ambient (benign) captures and 33 attack captures \cite{Verma2024RoadDataset}. The attacks in ROAD range from easy-to-detect to highly stealthy. Specifically, the dataset includes random fuzzing floods, targeted fabrication attacks that inject malicious data for specific CAN IDs. All CAN frames are timestamped and labeled (with the help of provided metadata) as either \emph{normal} or as a particular attack type, enabling supervised training and evaluation of intrusion detection models. %{Attack types considered include fabrication attacks (inserting fake CAN frames on the bus), fuzzing attacks (high-volume injection of random CAN frames to flood the bus), etc.}

\subsection{Dataset}

We use the ROAD CAN IDS dataset~\cite{Verma2024RoadDataset}, a high-fidelity collection of CAN bus traffic that includes both normal driving behavior and a diverse set of verified attack scenarios. The dataset comprises over 3.5 hours of CAN traffic collected from a real vehicle under dynamometer testing conditions, including 12 ambient (benign) captures and 33 attack captures. The attack scenarios span a range of difficulty, from easily detectable behaviors to more stealthy manipulations. In particular, ROAD includes random fuzzing floods as well as targeted fabrication attacks that inject malicious data for specific CAN IDs. All CAN frames are timestamped and labeled, with the aid of accompanying metadata, as either \emph{normal} or as a specific attack type, enabling supervised training and evaluation of intrusion detection models. 
The evaluated attacks include fuzzing attacks (FA) and fabrication attacks such as
max engine coolant temperature (MECTA), max speedometer (MSA), reverse light off (RLOFFA),
reverse light on (RLONA), and correlated signal (CSA) attacks.~\autoref{table:road_dataset_stats} summarizes the types of attacks used in this study.
\input{Tables/table_stats}

% \subsection{Preprocessing}\label{sub:prep}
% {CAN logs were parsed using a custom Python script}, which extracted the following fields from each frame: timestamp, CAN ID, and the data field payload. The timestamp was normalized to begin from zero. The CAN ID was converted from hexadecimal to decimal {\R Ans: followed UK paper}, and the data field was padded with leading zeros to ensure it contained 16 hexadecimal characters (8 bytes). The data bytes (D0–D7) were extracted from the 16-character hex string and converted to decimal. 

% Labels were added by comparing each frame with the known injected data rows (1 = malicious, 0 = benign). The final structured dataset was saved as CSV files. {\B Each frame was treated independently, using the CAN ID, DLC, and payload bytes (D0–D7) as features and the flag (0 or 1) as the label, without relying on temporal context.}

\subsection{Preprocessing}\label{sub:prep}
CAN logs were parsed using a custom Python script that extracted the %timestamp, CAN ID, and 
data field payload from each frame. 
%The timestamp was normalized to start from zero. The CAN ID was converted from hexadecimal to decimal, and 
The data field was padded with leading zeros to ensure a fixed length of 16 hexadecimal characters corresponding to eight bytes. The payload bytes (D0–D7) were then extracted from the hexadecimal string and converted to decimal values.

Labels were assigned by matching each frame against the known injected attack data, with frames labeled as benign (0) or malicious (1). The resulting structured dataset was stored in CSV format. Each frame was treated independently during preprocessing, using payload bytes (D0–D7) as features and the binary flag as the label, without incorporating temporal context or aggregation.

\subsection{Domain Constraints}

To ensure that adversarial examples remain valid CAN frames, we enforce protocol-level domain constraints during attack generation. Prior work has noted that many  AML studies overlook such constraints in networked systems, resulting in perturbations that may be effective in theory but invalid on the wire~\cite{mbow2021evaluating,li2020survey}. In classical CAN, each frame is defined by an ID, a DLC, and up to eight data bytes. In practice, ECUs transmit and accept only a fixed set of IDs configured by the manufacturer, and DLC values are determined by firmware and hardware constraints. Modifying these fields would therefore produce frames that are trivially rejected or never generated in real vehicles.

Under our threat model, the adversary operates exclusively on the payload bytes of existing frames. The CAN ID and DLC fields are held fixed to preserve message routing and length, while the eight data bytes (D0–D7) constitute the only mutable components. Accordingly, perturbations are restricted to payload-byte feature columns, ensuring that gradients are applied only to CAN data bytes. 
% while ID and DLC remain unchanged
%Dduring FGSM, BIM, and PGD attacks, gradients are applied exclusively to these payload entries, and all other features retain their original values at every attack step.

After each update, perturbed payload values are clipped to the valid CAN byte range [0, 255] and converted back to integer values prior to evaluation. By constraining the attacker in this manner, we emulate a realistic CAN-bus adversary who reuses legitimate IDs and timings while manipulating only the data field. This design choice is consistent with established in-vehicle threat models and recent CAN IDS studies~\cite{aloraini2024adversarial,Verma2024RoadDataset,mbow2021evaluating,li2020survey}.

\subsection{Adversarial Attack Methods}

To evaluate the robustness of the proposed IDS against adaptive adversaries, we generate adversarial examples using established gradient-based techniques. We leverage the Adversarial Robustness Toolbox (ART)~\cite{nicolae2018art} to craft attacks that introduce subtle, targeted manipulations of CAN payload bytes with the goal of inducing misclassification. An overview of the adversarial workflow is illustrated in \autoref{fig:road_workflow_fn_fp_mcc}. The following attack methods are considered in our experiments:
\input{Figures/figure_workflow}
\begin{itemize}[left=0pt]
\item \textbf{FGSM}~\cite{goodfellow2015explaining}: A single-step white-box attack that computes the gradient of the IDS loss with respect to the input features and perturbs the input in the direction of the sign of the gradient. We apply FGSM using a small perturbation budget $\epsilon$, adding an $\epsilon$-scaled sign of the gradient to each payload feature. This produces adversarial CAN frames with minimal changes to data values that can nonetheless cause the IDS to misclassify individual frames.

\item \textbf{BIM}~\cite{Kurakin2016BIM}: An iterative extension of FGSM, also referred to as I-FGSM. BIM applies multiple small gradient-based updates, each scaled by a fraction of $\epsilon$, while clipping the perturbed input after every step to preserve validity. This iterative refinement results in stronger adversarial examples that remain within the allowed perturbation bounds.

\item \textbf{PGD}~\cite{Madry2018PGD}: A multi-step attack that generalizes BIM by incorporating random initialization within the allowed perturbation region. PGD performs a sequence of gradient-based updates and projects the perturbed input back into the $\epsilon$-ball around the original sample after each step. Often regarded as a strong first-order adversary, PGD approximates a worst-case attack against a given model by generating minimally perturbed CAN frames that maximize prediction error under full IDS knowledge.
\end{itemize}

In this work, FGSM, BIM, and PGD attacks are generated using a differentiable DNN surrogate and the resulting adversarial samples are then reused to evaluate the robustness of shallow models in a transfer-based setting under identical protocol-constrained perturbations consistent with Pacheco \textit{et al.}~\cite{pacheco2021adversarial}. 
Also, we evaluate IDS robustness under bounded payload perturbations applied to frames labeled benign or malicious in the ROAD dataset without claiming semantic or physical class preservation. Following the use of $L_\infty$-bounded perturbations in Carlini et al.~\cite{carlini2017towards}, the perturbation budget 
$\epsilon$ enforces a per-byte bound with $\epsilon\in\{1,5\}$ in our experiments, yielding small, valid-range numerical changes that constrain perturbation magnitude for both benign and attack data, but do not guarantee semantic invariance, and thus this setup enables the evaluation of false alarms and missed detections under adversarial conditions.

\subsection{Evaluation Metrics}

We evaluate adversarial effectiveness and classifier performance using two complementary metrics: the attack success rate (ASR)~\cite{li2019nattack} and the Matthews correlation coefficient (MCC)~\cite{chicco2020advantages}. ASR quantifies how often crafted adversarial samples successfully induce misclassification by the IDS. Formally, let $N_{\text{[adv]}}$ denote the number of adversarially perturbed samples and $N_{\text{[succ]}}$ the number of those samples that are misclassified by the IDS. ASR is then defined as $\mathrm{ASR} = \tfrac{N_{\text{[succ]}}}{N_{\text{[adv]}}}.$

In the IDS setting considered here, ASR captures two distinct adversarial objectives: inducing benign frames to be flagged as malicious (false positives) and causing malicious frames to be misclassified as normal (false negatives). Expressed in terms of confusion-matrix entries, the ASR for benign inputs, corresponding to adversarially induced false alarms is $\mathrm{ASR}_{[FP]}$,  while for malicious inputs, it is $\mathrm{ASR}_{[FN]}$ expressed as\\[-3ex]
$${
\mathrm{ASR}_{[FP]} = \tfrac{\mathrm{FP}_{[adv]}}{\mathrm{TN}_{[adv]} + \mathrm{FP}_{[adv]}},\;
\mathrm{ASR}_{[FN]} = \tfrac{\mathrm{FN}_{[adv]}}{\mathrm{TP}_{[adv]} + \mathrm{FN}_{[adv]}}}\\[-.5ex]
$$
respectively. Although these expressions resemble the standard false negative rate and false positive rate, they are computed exclusively over adversarially perturbed samples. As such, they directly quantify attacker success in terms of missed attacks and induced false alarms. Higher ASR values indicate more effective adversarial attacks and reduced IDS robustness.

In addition to ASR, we report MCC to assess overall classification performance under class imbalance. FP and FN are computed from evaluations on benign-only and malicious-only subsets, respectively, to isolate false-alarm and miss-detection behavior. In contrast, MCC is evaluated on the full test set containing both benign and malicious frames, where joint classification performance under class imbalance can be meaningfully assessed.
The ROAD dataset contains approximately 1.5 million benign frames and 50 thousand malicious frames, resulting in a strongly imbalanced class distribution of about 3.2\% malicious traffic~\cite{Verma2024RoadDataset}. Under such conditions, commonly used metrics such as the F1 score can overestimate performance by favoring the majority class~\cite{aloraini2024adversarial}. MCC is well suited to this setting because it incorporates all four confusion-matrix components and provides a balanced summary of binary classification quality. Given the counts $\mathrm{TP}$, $\mathrm{TN}$, $\mathrm{FP}$, and $\mathrm{FN}$, MCC is defined as\\[-2ex]
\[
\mathrm{MCC} = \tfrac{\mathrm{TP}\cdot\mathrm{TN} - \mathrm{FP}\cdot\mathrm{FN}}
{\sqrt{(\mathrm{TP}+\mathrm{FP})(\mathrm{TP}+\mathrm{FN})(\mathrm{TN}+\mathrm{FP})(\mathrm{TN}+\mathrm{FN})}}.\\[-.5ex]
\]
MCC ranges from $-1$ (total disagreement) to $0$ (no better than random) and up to $+1$ (perfect prediction). We report MCC alongside  FP/FN counts and their respective ASR of benign-only and malicious-only subsets to provide a robust view of IDS performance that penalizes both false alarms and missed-detections within a single scalar measure.

%% file: Figures/figure_CAN.tex
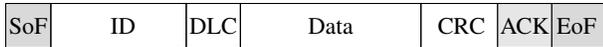
\begin{figure}[htbp]
\centering
\begin{tikzpicture}[scale=0.6, font=\small, every node/.style={align=center}]
  % overall height
  \def\H{1}

  % --- SOF ---
  \fill[gray!30] (0,0) rectangle (1,\H);
  \draw (0,0) rectangle (1,\H);
  \node at (0.5,0.5*\H) {SoF};

  % --- Identifier (slightly shortened) ---
  \draw (1,0) rectangle (4,\H);
  \node at (2.6,0.5*\H) {ID};

  % --- DLC ---
  \draw (4,0) rectangle (5.2,\H);
  \node at (4.6,0.5*\H) {DLC};

  % --- Data (shortened a bit more) ---
  \draw (5.2,0) rectangle (9.2,\H);
  \node at (7.2,0.5*\H) {Data};

  % --- CRC ---
  \draw (9.2,0) rectangle (10.9,\H);
  \node at (10.1,0.5*\H) {CRC};

  % --- ACK ---
  \fill[gray!20] (10.9,0) rectangle (12.1,\H);
  \draw (10.9,0) rectangle (12.1,\H);
  \node at (11.5,0.5*\H) {ACK};

  % --- EOF ---
  \fill[gray!30] (12.1,0) rectangle (13.3,\H);
  \draw (12.1,0) rectangle (13.3,\H);
  \node at (12.65,0.5*\H) {EoF};
\end{tikzpicture}
\caption{CAN data frame structure.}
\label{fig:can_frame}
\end{figure}

%% file: Tables/table_stats.tex
\begin{table}[!hbt]
\setlength{\tabcolsep}{2.5pt} % default is 6pt
\centering
\caption{Statistical Breakdown of Attacks in the ROAD Dataset.}
\vspace{1mm}
\begin{tabular}{|l|r|r|r|}
\hline
\textbf{Attack Type} & \textbf{Benign} & \textbf{Malicious} & \textbf{Total} \\
\hline
FA  & 87,907 & 1,061 & 88,968 \\
\hline
MECTA & 57,932 & 88 & 58,020 \\
\hline
MSA & 520,216 & 17,221 & 537,437 \\
\hline
RLOFFA  & 180,357 & 10,605 & 190,962 \\
\hline
RLONA & 480,021 & 15,195 & 495,216 \\
\hline
CSA & 175,412 & 5,493 & 180,905 \\
\hline
\textbf{Total} & \textbf{1,501,845} & \textbf{49,663} & \textbf{1,551,508} \\
\hline
\end{tabular}
\label{table:road_dataset_stats}
\end{table}

% \begin{table}{!hbt}
% \setlength{\tabcolsep}{2.5pt}
% \centering
% \caption{Statistical Breakdown of Attacks in the ROAD Dataset.}
% \label{table:road_dataset_stats}
% \begin{tabular}{|l|r|r|r|}
% \toprule
% Attack Type & Benign & Malicious & Total \\
% \midrule
% CSA & 175,412 & 5,493 & 180,905 \\
% FA & 87,907 & 1,061 & 88,968 \\
% MECTA & 57,932 & 88 & 58,020 \\
% MSA & 520,216 & 17,221 & 537,437 \\
% RLOFFA & 180,357 & 10,605 & 190,962 \\
% RLONA & 480,021 & 15,195 & 495,216 \\
% \textbf{Total} & \textbf{1,501,845} & \textbf{49,663} & \textbf{1,551,508} \\
% \bottomrule
% \end{tabular}
% \end{table}

%% file: Figures/figure_workflow.tex
\tikzset{
  block/.style = {
      rectangle, rounded corners, draw, thick,
      align=center, minimum width=3cm, minimum height=1.2cm,  % larger
      font=\bfseries\normalsize                             % bold + larger text
  },
  smallblock/.style = {
      rectangle, draw, thick, align=center,
      minimum width=1.6cm, minimum height=0.9cm,      font=\bfseries\small                                  % bold small text
  },
  decision/.style = {
      diamond, draw, thick, align=center,
      aspect=2, inner sep=2pt,
      font=\bfseries\normalsize                              % bold decision text
  },
  shufflebadge/.style={
   rounded corners=4pt,
   inner sep=4pt,
   % top color=blue!10,
   % bottom color=blue!30,
   draw=black
 },
  arrow/.style = {->, >=Stealth, thick}
}

\begin{figure}[t]
\centering
\resizebox{\columnwidth}{!}{%
\begin{tikzpicture}[node distance=1.5cm and 1cm]
font=\bfseries\normalsize
% Left input stack
\node (fa) [input] {FA};
\node (mecta) [input, below=0.2cm of fa] {MECTA};
\node (msa) [input, below=0.2cm of mecta] {MSA};
\node (rloffa) [input, below=0.2cm of msa] {RLOFFA};
\node (rlona) [input, below=0.2cm of rloffa] {RLONA};
\node (csa) [input, below=0.2cm of rlona] {CSA};
% Outer box for attack datasets

\node (normal) [input, below=0.8cm of csa,   fill=green!10] {Normal};
\node (attack) [input, below=0.2cm of normal,  fill=blue!10] {Attack};
\node(road)[draw,
      rounded corners,
      dashed, draw=blue,
      inner sep=18pt,
      fit=(fa) (attack),
      label={[font=\bfseries]left:}] {};
%\node (attack) [input, below=0.2cm of normal, draw=red, dashed, fill=blue!10] {Attack};
\node(mal)[draw,
      rounded corners,
      dashed, 
      inner sep=8pt, draw=red,
      fit=(fa) (csa), line width=1.5pt,
      label={[font=\bfseries]left:}] {};

\node(noratt)[draw,
      rounded corners,
      dashed, line width=1.5pt,
      inner sep=8pt, draw=Dred,
      fit=(normal) (attack),
      label={[font=\bfseries]left:}] {};

% IDS pipeline box
\node (pre) [process, right=1.5cm of rloffa, minimum width=3cm, yshift=-0.9cm ] {Pre-processing {\large \faTools} };
\node (train) [process, right=3.2cm of mecta,fill=gray!5, minimum width=3cm, align=center] {{\Large\dr \faCogs} \\ IDS Training\\ and Testing};
\node (pred) [shufflebadge, fill=gray!15, right=1.2cm of train, minimum width=2.7cm,, align=center] {{\Large\dr\faRandom }\\ IDS\\ {\large Predictions}};

% Adversarial methods
\node (fgsm) [adversary, below=2.3cm of csa
, xshift=3.7cm] {FGSM};
\node (bim) [adversary, right=0.3cm of fgsm] {BIM};
\node (pgd) [adversary, right=0.3cm of bim] {PGD};

\node(adv)[draw,
      rounded corners,
      dashed, 
      inner sep=8pt, draw=dred,
      fit=(fgsm) (pgd), line width=1.5pt,
      label={[font=\bfseries]left:}] {};

% Adversarial Frame Generator
\node (gen) [process, above=0.9cm of bim,fill=gray!5, minimum width=3cm, align=center] {{\Large\dr\faBug}\\Adversarial IVN \\ Frame Generation};

% Constraint decision
\node (cond) [decision, right=0.9 cm of gen,fill=gray!7, xshift=0 cm] {Constraint\\ Compliant?};

% Outputs
\node (benign) [output, right=2cm of pred, yshift=1.2cm, minimum width=2.5cm, fill=red!10] { FP};
\node (malicious) [output, right=2cm of pred, minimum width=2.5cm,fill=red!10] {FN};
\node (fn) [output, below=1cm of malicious, minimum width=2.5cm, fill=red!30] {ASR};
\node (fp) [output, below=1cm of fn, minimum width=2.5cm, fill=blue!30] {MCC};
\node (discard) [output, right=1.4cm of cond, minimum width=2.5cm,fill=yellow!30] {Discarded};

% Arrows
\draw [arrow] (road.east) --  (pre.west);
\draw [arrow] (mal.west) -- ++(-0.7,0) |- (noratt.west);
\draw [arrow] (pre) -- (train);
\draw [arrow] (train) -- (pred);
\draw [arrow] (pre.south) -- (gen.north);
\draw [arrow] (adv.north)  -- ++(0,0.5) -| (gen.south);
%\draw [arrow] (bim.north) -- ++(0,0.5) -| (gen.south);
%\draw [arrow] (pgd.north)  -- ++(0,0.5) -| (gen.south);
\draw [arrow] (gen) -- (cond);
\draw [arrow] (cond.north)  --node[right, xshift=1pt]{Yes {\large \faCarSide}} (pred.south);
\draw [arrow] (cond.east) --node[above, xshift=1pt]{No {\R \faBan}} (discard);
\draw [arrow] (pred.east) -- ++(1,0) |- (benign.west);
\draw [arrow] (pred.east) -- ++(1,0) |- (malicious.west);
\draw [arrow] (pred.east) -- ++(1,0) |- (fn.west);
\draw [arrow] (pred.east) -- ++(1,0) |- (fp.west);

\end{tikzpicture}
}
\caption{Workflow for IDS training, adversarial IVN frame generation, and evaluation of benign and adversarial predictions with FN, FP, and MCC.}
\label{fig:road_workflow_fn_fp_mcc}
\end{figure}
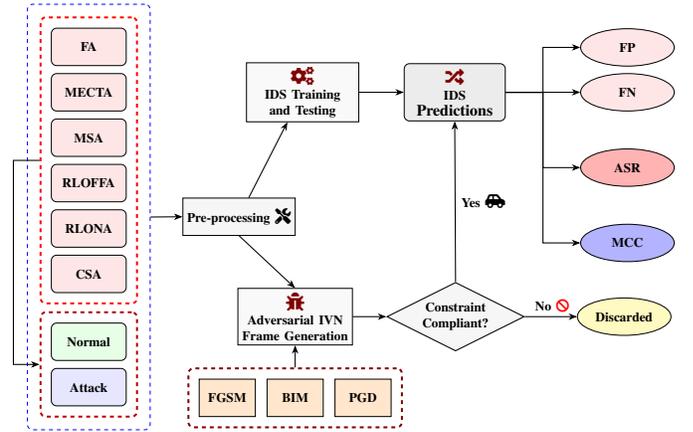

%% file: 5_results.tex
\section{Results}
{
% \subsection{Overview of Evaluation}
% {\B \autoref{table:road_dataset_stats} shows that the IDS classifiers were trained on a merged dataset containing six subsets, FA + five fabrication attacks (MECTA, MSA, RLOFFA, RLONA and CSA), totaling 1,540,441 samples obtained after the preprocessing steps. This dataset includes 1,490,778 benign and 49,663 malicious samples. We use a 70/30 split, resulting in 1,078,308 samples for training and 462,133 samples for testing (96.8\% benign, 3.2\% malicious), with class proportions preserved under random shuffling.}

%  %{\R What did we learn from this?}
% \autoref{table:ids_performance} summarizes the performance of each IDS {\B under $30\%$ test samples} by showing FPs, FNs and MCC values. The results show that the shallow models achieve strong baseline detection with MCC values close to $0.90$. {\B On the other hand, DNN attains a slightly lower MCC ($\approx 0.84$), but with fewer FPs but far more FNs, indicating a more conservative benign-class decision boundary.} This benign baseline establishes that any degradation observed under adversarial perturbations is attributable to the attack, rather than to poor nominal IDS performance.

% \autoref{tab:fa} to \autoref{tab:fn_rlon} present the results evaluating FPs, FNs, ASR, and MCC across all models under different attack scenarios. 

% However, \autoref{tab:csa} doesn't produce FNs and therefore displays the results finding F1-scores instead of MCC.

\subsection{Overview of Evaluation}
% Total 1,501,845 49,663 1,551,508
\autoref{table:road_dataset_stats} summarizes the data used for training and evaluation. The IDS classifiers are trained on a merged dataset comprising six subsets including a fuzzing attack and five fabrication attacks for a total of 1,551,508 samples after preprocessing. 
This dataset contains 1,501,845 benign samples and 49,663 malicious samples. 
We apply a 70/30 train–test split, resulting in 1,086,055 samples for training and 465,453 samples for testing, with class proportions preserved under random shuffling (96.8\% benign and 3.2\% malicious).
% Train samples: 1086055 Test samples: 465453 Total samples:  1551508

\autoref{table:ids_performance} reports the baseline performance of each IDS on the 30\% test set, including false positives (FPs), false negatives (FNs), and MCC values. The results indicate that shallow models achieve strong nominal detection performance, with MCC values close to 0.91. In contrast, the DNN attains a slightly lower MCC (approximately 0.85), exhibiting fewer false positives but substantially more false negatives, which suggests a more conservative decision boundary favoring the benign class. This baseline evaluation establishes that any performance degradation observed under adversarial perturbations can be attributed to the attacks themselves rather than to inadequate nominal IDS performance.

\autoref{tab:fa} through \autoref{tab:rlona} present results for adversarial evaluation, reporting FPs, FNs, ASR, and MCC across all models and attack scenarios. For the CSA subset, shown in \autoref{tab:csa}, no false negatives are produced; accordingly, results for this case are reported using F1 scores since MCC is zero when false negatives are not produced.
\input{Tables/table_performance_metrics}
\input{Tables/table_fa}
\input{Tables/table_mecta}
\input{Tables/table_msa}
\subsection{Benign Performance and False-Alarm Robustness}
Across all attack methods (FGSM, BIM, and PGD) and perturbation budgets ($\epsilon = 1, 5$), the IDS models generally maintain stable benign classification performance, as reflected by the MCC values that remain close to their nominal baselines. With the exception of PGD at $\epsilon = 5$, which leads to noticeable degradation, shallow models largely preserve their ability to correctly classify benign traffic under adversarial perturbations. On the other hand, the DNN exhibits significantly lower ASR accurately capturing benign frames than other models in benign settings under adversarial conditions.
%failing to correctly classify benign frames and 

%except MECTA where it fails to capture the malicious frames. 
This behavior suggests that the DNN demonstrates superior robustness to adversarial perturbations on benign traffic, limiting false-alarm induction compared to shallow models.
\input{Tables/table_rloffa}

\input{Tables/table_rlona}
\input{Tables/table_csa}
% \subsection{Adversarial Performance and Missed-Attack Vulnerability}
% Results for adversarial perturbations applied to malicious samples, reported in \autoref{tab:fa} through \autoref{tab:csa}, reveal significant missed-attack vulnerabilities across the evaluated IDS models. Under FGSM, BIM, and PGD attacks, the DNN consistently exhibits an attack success rate (ASR) of 100\% across all attack types, completely failing to detect adversarially manipulated malicious frames and yielding zero MCC in all cases. Similarly, XGB experiences severe degradation, with ASR values exceeding 95\% in most scenarios. These results are consistent with prior findings showing that gradient-based adversarial methods can effectively exploit the differentiable decision boundaries of learning-based models~\cite{goodfellow2015explaining, carlini2017towards, nicolae2018art}. 

% In contrast, the ET model demonstrates comparatively higher robustness across most attack types, with notably lower ASR values, although it remains vulnerable in the RLOFFA scenario, where ASR approaches 20\%. For the FA dataset specifically, all models except the DNN exhibit relatively low ASR, indicating better resilience to missed attacks in this setting.
\subsection{Adversarial Performance and Missed-Attack Vulnerability}
Results for adversarial perturbations applied to malicious samples, summarized in \autoref{tab:fa} through \autoref{tab:csa}, reveal substantial missed-attack vulnerabilities across the evaluated IDS models. 
Under FGSM, BIM, and PGD attacks on MECTA, the DNN consistently exhibits an attack success rate (ASR) of 1.0, indicating a complete failure to detect adversarially manipulated malicious frames and resulting in non-positive MCC. The shallow models also yield close to zero MCC on MECTA, implying performance no better than random guessing.
Similarly, the shallow models suffer severe degradation for PGD at both perturbations compared to FGSM and BIM, significantly failing to capture malicious frames with ASR increasing from near-zero under FGSM/BIM to above 0.5 in many cases, reaching as high as 0.95 under same model.
These observations are consistent with prior work showing that gradient-based adversarial methods can effectively exploit the differentiable decision boundaries of learning-based models~\cite{goodfellow2015explaining, carlini2017towards, nicolae2018art}.

%While Extra Trees and Random Forest exhibit comparable and very low FN-ASR (≈0.01–0.02) under FGSM and BIM, Extra Trees degrades more gracefully under stronger PGD attacks—typically maintaining FN-ASR below ~0.5—whereas Random Forest frequently exceeds this level and in several cases approaches ~0.9, a trend that is consistent across datasets.

In contrast, the ET model demonstrates comparatively higher robustness across most attack types, achieving notably lower ASR values. Although ET and RF perform similarly under FGSM and BIM, ET shows more stable missed-attack robustness under PGD, with lower and more gradual ASR increases than RF across all scenarios. However, all the models including ET remain vulnerable in the RLOFFA scenario, where ASR exceeds 0.43. For the FA dataset specifically, all models exhibit relatively low ASR, indicating greater resilience to missed attacks in this setting.

%Although the DNN achieves superior benign robustness, it exhibits scenario-dependent catastrophic missed-attack behavior, reaching FN-ASR = 1.0 in certain settings, whereas the best shallow model (Extra Trees) maintains bounded FN-ASR (generally below ~0.6) across all evaluated scenarios, resulting in more consistent missed-attack robustness overall.

% \subsection{Gradient-Based and Protocol-Aware Vulnerabilities}
% The results demonstrate that gradient-based attacks (FGSM, BIM, PGD) effectively deceive 
% {\B models directly or indirectly exploitable via gradient-based perturbations, including DNN and XGB}, producing nearly total bypassing. This aligns with previous findings that adversarial perturbations produced via gradient sign methods can induce high-confidence misclassification in deep networks~\cite{goodfellow2015explaining, carlini2017towards}. 

% Conversely, protocol-aware CAN fabrication attacks that manipulate the semantic meaning of message payloads (e.g., MSA, RLOFFA, RLONA) strongly affected tree-based models. Although ET showed comparatively lower ASRs (21--26\% for several cases), DT and RF models often exceeded 90\% ASR. These results are consistent with earlier observations that ensemble averaging can mitigate but not eliminate the vulnerabilities in tree-based learners~\cite{papernot2016transferability}.

\subsection{Gradient-Based and Protocol-Aware Vulnerabilities}
The results show that gradient-based attacks (FGSM, BIM, and PGD) are highly effective against models that are directly or indirectly exploitable to gradient-based perturbations, including DNN leading to complete evasion in specific cases.
This behavior is consistent with prior findings demonstrating that gradient-sign–based perturbations can induce high-confidence misclassification in DL models~\cite{goodfellow2015explaining, carlini2017towards}.

In contrast, protocol-aware CAN fabrication attacks that manipulate the semantic content of payloads, such as MSA, RLOFFA, and RLONA, have a stronger impact on tree-based models. 
While ET typically attains lower ASR than the other shallow models, DT, RF and XGB can attain significantly higher ASR, even approach near-complete evasion in missed-detections isolated cases (especially for PGD), exceeding ASR of 0.90.
% all models including ET remain vulnerable in RLOFFA,  
These results align with earlier observations that ensemble-based methods can reduce, but not fully eliminate, adversarial vulnerabilities in tree-based learners~\cite{papernot2016transferability}.

% \subsection{Comparative Robustness Across Models}
% Overall, most models remained robust against benign perturbations (low FP ASR), yet vulnerable to adversarial manipulation of malicious samples. {\B Among the shallow models, } ET demonstrated the best robustness balance, minimal FPs and lower FNs. DNN exhibited the most severe degradation (100\% FN ASR across attacks), confirming the weakness of differentiable architectures in constrained IVN settings. These findings assert the necessity of adversarial evaluation for automotive IDS and support the use of hybrid training to improve flexibility~\cite{nicolae2018art, apruzzese2022adversarial}.

% The CSA did not yield any FN results during adversarial evaluation. This is a direct consequence of the protocol-level constraints enforced during adversarial sample generation~\cite{mbow2021evaluating,li2020survey, wasicek2017context}. We allowed perturbations only on the eight data bytes, while the ID and DLC fields were kept fixed and all modified bytes were clipped to the valid range $[0,255]$. Any frame violating these constraints was discarded to preserve CAN-frame validity. Under this masking and clipping scheme, CSA frames were excluded because the attack pattern requires coordinated changes across multiple correlated signals, which conflicts with the imposed fixed-field constraints. As a result, no valid adversarial CSA samples reached the IDS, and no FN values are reported for this attack type.

\subsection{Comparative Robustness Across Models}
Overall, most models remain robust to benign perturbations, as reflected by low false-positive ASR, but are substantially more vulnerable to adversarial manipulation of malicious samples. 
Although the DNN achieves superior benign robustness, it exhibits scenario-dependent catastrophic missed-attack behavior, reaching missed-attack ASR = 1.0 in adversarial settings, whereas ET maintains bounded in ASR (generally below ~0.6) across all evaluated scenarios, resulting in more consistent missed-attack robustness overall.
Thus, ET demonstrates the most favorable robustness trade-off, exhibiting minimal false alarms and comparatively lower missed-attack rates.
%In contrast, the DNN shows the most severe degradation, with a 100\% false-negative ASR across attack types, highlighting the fragility of differentiable architectures under constrained in-vehicle network settings. 
These findings underscore the importance of adversarial evaluation for automotive IDS and motivate the use of robustness-oriented training strategies to improve resilience~\cite{nicolae2018art, apruzzese2022adversarial}.

The CSA attack does not produce any false negatives during adversarial evaluation. This outcome follows directly from the protocol-level constraints imposed during adversarial sample generation~\cite{mbow2021evaluating, li2020survey, wasicek2017context}. Perturbations are restricted to the eight data bytes, while CAN ID and DLC fields remain fixed and all modified values are clipped to the valid range $[0,255]$. Frames violating these constraints are discarded to preserve CAN-frame validity. 
%Under this masking and clipping scheme, CSA samples are excluded because the attack pattern requires coordinated modifications across multiple correlated signals, which conflicts with the enforced fixed-field constraints. 
Consequently, no valid adversarial CSA samples reach the IDS, and no false-negative results are reported for this attack type.
% Furthermore, prior research indicates that multi-signal attacks like CSA are inherently easier to detect~\cite{aloraini2024adversarial, wasicek2017context}. When all correlated signals (wheel speeds) are modified together, the resulting frame deviates sharply from expected inter-signal relationships, producing conspicuous anomalies~\cite{9521818}.
% For this reason, even if CSA perturbations were permitted, they would likely be measured with near-perfect scores, consistent with previous findings that reported 100\% detection for these types of manipulations.
% Therefore, the lack of FN values for CSA reflects both the protocol-level validity constraints and the natural detectability of multi-signal attacks in IVN traffic.

}

%% file: Tables/table_performance_metrics.tex
\begin{table}[!t]
\setlength{\tabcolsep}{2.5pt} % default is 6pt
\centering
\caption{Performance Metrics  on  Test Samples Under Benign Settings.}
\label{table:ids_performance}
\begin{tabular}{|c|c|c|c|c|c|}
\hline
\thead{Model} &
\thead{Benign} &
\thead{Malicious} &
\thead{FP} &
\thead{FN} &
\thead{MCC} \\
\hline
DT & \multirow{5}{*}{450,554} & \multirow{5}{*}{14,899} & 941 & 1,668 & 0.908 \\
RF &  &  & 951 & 1,659 & 0.908 \\
ET &  &  & 929 & 1,683 & 0.907 \\
XGB &  &  & 966 & 1,638 & 0.908 \\
DNN &  &  & 384 & 3,781 & 0.845 \\
\hline
\end{tabular}
\end{table}

%% file: Tables/table_fa.tex
\begin{table}[!t]
\scriptsize
\centering
\scriptsize
\setlength{\tabcolsep}{2.5pt} % default is 6pt
\caption{Comparative analysis -- FA.}
\label{tab:fa}
\renewcommand\arraystretch{1}
\begin{tabular}{|c|ccc|c|c|cc|cc|c|}
\hline
 &
\multicolumn{3}{c|}{\textbf{Benign}} &
 & &
\multicolumn{5}{c|}{\textbf{Adversarial}}\\
\cline{2-4}\cline{7-11}
\textbf{Model} &
\textbf{FP} & \textbf{FN} & \textbf{MCC} &
\textbf{Attack} & \boldmath$\epsilon$ &
\textbf{FP} & \textbf{ASR} & \textbf{FN} & \textbf{ASR} & \textbf{MCC}\\

\hline
\multirow{6}{*}{DT} & 22 & 0 & 0.99 & FGSM & 1 & 67 & 0.001 & 0 & 0.0 & 0.838\\
 & 22 & 0 & 0.99 & FGSM & 5 & 67 & 0.001 & 0 & 0.0 & 0.789\\
 & 22 & 0 & 0.99 & BIM & 1 & 67 & 0.001 & 0 & 0.0 & 0.833\\
 & 22 & 0 & 0.99 & BIM & 5 & 67 & 0.001 & 0 & 0.0 & 0.833\\
 & 22 & 0 & 0.99 & PGD & 1 & 188 & 0.003 & 84 & 0.134 & 0.777\\
 & 22 & 0 & 0.99 & PGD & 5 & 294 & 0.005 & 192 & 0.306 & 0.638\\
\hline
\multirow{6}{*}{RF} & 20 & 0 & 0.989 & FGSM & 1 & 8 & 0.0 & 0 & 0.0 & 0.998\\
 & 20 & 0 & 0.989 & FGSM & 5 & 8 & 0.0 & 0 & 0.0 & 0.994\\
 & 20 & 0 & 0.989 & BIM & 1 & 8 & 0.0 & 0 & 0.0 & 0.998\\
 & 20 & 0 & 0.989 & BIM & 5 & 8 & 0.0 & 0 & 0.0 & 0.998\\
 & 20 & 0 & 0.989 & PGD & 1 & 11 & 0.0 & 84 & 0.134 & 0.978\\
 & 20 & 0 & 0.989 & PGD & 5 & 38 & 0.001 & 207 & 0.33 & 0.812\\
\hline
\multirow{6}{*}{ET} & 18 & 0 & 0.992 & FGSM & 1 & 8 & 0.0 & 0 & 0.0 & 1.0\\
 & 18 & 0 & 0.992 & FGSM & 5 & 8 & 0.0 & 0 & 0.0 & 1.0\\
 & 18 & 0 & 0.992 & BIM & 1 & 8 & 0.0 & 0 & 0.0 & 1.0\\
 & 18 & 0 & 0.992 & BIM & 5 & 8 & 0.0 & 0 & 0.0 & 1.0\\
 & 18 & 0 & 0.992 & PGD & 1 & 4 & 0.0 & 0 & 0.0 & 1.0\\
 & 18 & 0 & 0.992 & PGD & 5 & 31 & 0.001 & 0 & 0.0 & 0.995\\
\hline
\multirow{6}{*}{XGB} & 22 & 0 & 0.99 & FGSM & 1 & 69 & 0.001 & 0 & 0.0 & 0.994\\
 & 22 & 0 & 0.99 & FGSM & 5 & 69 & 0.001 & 0 & 0.0 & 0.948\\
 & 22 & 0 & 0.99 & BIM & 1 & 69 & 0.001 & 0 & 0.0 & 0.983\\
 & 22 & 0 & 0.99 & BIM & 5 & 69 & 0.001 & 0 & 0.0 & 0.983\\
 & 22 & 0 & 0.99 & PGD & 1 & 34 & 0.001 & 312 & 0.499 & 0.898\\
 & 22 & 0 & 0.99 & PGD & 5 & 90 & 0.001 & 321 & 0.512 & 0.742\\
\hline
\multirow{6}{*}{DNN} & 19 & 24 & 0.978 & FGSM & 1 & 6 & 0.0 & 24 & 0.038 & 0.992\\
 & 19 & 24 & 0.978 & FGSM & 5 & 6 & 0.0 & 24 & 0.038 & 0.803\\
 & 19 & 24 & 0.978 & BIM & 1 & 6 & 0.0 & 24 & 0.038 & 0.992\\
 & 19 & 24 & 0.978 & BIM & 5 & 6 & 0.0 & 24 & 0.038 & 0.992\\
 & 19 & 24 & 0.978 & PGD & 1 & 6 & 0.0 & 26 & 0.042 & 0.99\\
 & 19 & 24 & 0.978 & PGD & 5 & 43 & 0.001 & 26 & 0.041 & 0.99\\
\hline
\end{tabular}
\end{table}

%% file: Tables/table_mecta.tex
\begin{table}[!t]
\scriptsize
\setlength{\tabcolsep}{2.5pt} % default is 6pt
\centering
\caption{Comparative analysis -- MECTA.}
\label{tab:mecta}
\renewcommand\arraystretch{1}
\begin{tabular}{|c|ccc|c|c|cc|cc|c|}
\hline
 &
\multicolumn{3}{c|}{\textbf{Benign}} &
 & &
\multicolumn{5}{c|}{\textbf{Adversarial}}\\
\cline{2-4}\cline{7-11}
\textbf{Model} &
\textbf{FP} & \textbf{FN} & \textbf{MCC} &
\textbf{Attack} & \boldmath$\epsilon$ &
\textbf{FP} & {\bf ASR} & \textbf{FN} & \textbf{ASR} & \textbf{MCC}\\
\hline
\multirow{6}{*}{DT} & 17 & 26 & 0.748 & FGSM & 1 & 14 & 0.0 & 4 & 0.667 & -0.0\\
 & 17 & 26 & 0.748 & FGSM & 5 & 14 & 0.0 & 4 & 0.667 & -0.0\\
 & 17 & 26 & 0.748 & BIM & 1 & 14 & 0.0 & 4 & 0.667 & -0.0\\
 & 17 & 26 & 0.748 & BIM & 5 & 14 & 0.0 & 4 & 0.667 & -0.0\\
 & 17 & 26 & 0.748 & PGD & 1 & 159 & 0.004 & 3 & 0.5 & 0.051\\
 & 17 & 26 & 0.748 & PGD & 5 & 279 & 0.007 & 4 & 0.667 & -0.001\\
\hline
\multirow{6}{*}{RF} & 16 & 27 & 0.748 & FGSM & 1 & 13 & 0.0 & 4 & 0.667 & 0.0\\
 & 16 & 27 & 0.748 & FGSM & 5 & 13 & 0.0 & 4 & 0.667 & 0.0\\
 & 16 & 27 & 0.748 & BIM & 1 & 13 & 0.0 & 4 & 0.667 & 0.0\\
 & 16 & 27 & 0.748 & BIM & 5 & 13 & 0.0 & 4 & 0.667 & 0.0\\
 & 16 & 27 & 0.748 & PGD & 1 & 26 & 0.001 & 3 & 0.5 & 0.085\\
 & 16 & 27 & 0.748 & PGD & 5 & 11 & 0.0 & 4 & 0.667 & -0.0\\
\hline
\multirow{6}{*}{ET} & 16 & 27 & 0.733 & FGSM & 1 & 13 & 0.0 & 4 & 0.667 & -0.0\\
 & 16 & 27 & 0.733 & FGSM & 5 & 13 & 0.0 & 4 & 0.667 & 0.0\\
 & 16 & 27 & 0.733 & BIM & 1 & 13 & 0.0 & 4 & 0.667 & -0.0\\
 & 16 & 27 & 0.733 & BIM & 5 & 13 & 0.0 & 4 & 0.667 & -0.0\\
 & 16 & 27 & 0.733 & PGD & 1 & 24 & 0.001 & 3 & 0.5 & 0.091\\
 & 16 & 27 & 0.733 & PGD & 5 & 50 & 0.001 & 4 & 0.667 & -0.0\\
\hline
\multirow{6}{*}{XGB} & 16 & 27 & 0.733 & FGSM & 1 & 13 & 0.0 & 4 & 0.667 & -0.0\\
 & 16 & 27 & 0.733 & FGSM & 5 & 13 & 0.0 & 4 & 0.667 & -0.0\\
 & 16 & 27 & 0.733 & BIM & 1 & 13 & 0.0 & 4 & 0.667 & -0.0\\
 & 16 & 27 & 0.733 & BIM & 5 & 13 & 0.0 & 4 & 0.667 & -0.0\\
 & 16 & 27 & 0.733 & PGD & 1 & 13 & 0.0 & 6 & 1.0 & -0.0\\
 & 16 & 27 & 0.733 & PGD & 5 & 60 & 0.001 & 5 & 0.833 & -0.0\\
\hline
\multirow{6}{*}{DNN} & 0 & 88 & -0.0 & FGSM & 1 & 0 & 0.0 & 6 & 1.0 & 0.0\\
 & 0 & 88 & -0.0 & FGSM & 5 & 0 & 0.0 & 6 & 1.0 & -0.001\\
 & 0 & 88 & -0.0 & BIM & 1 & 0 & 0.0 & 6 & 1.0 & 0.0\\
 & 0 & 88 & -0.0 & BIM & 5 & 0 & 0.0 & 6 & 1.0 & 0.0\\
 & 0 & 88 & -0.0 & PGD & 1 & 7 & 0.0 & 6 & 1.0 & 0.0\\
 & 0 & 88 & -0.0 & PGD & 5 & 746 & 0.018 & 6 & 1.0 & -0.001\\
\hline
\end{tabular}
\end{table}

%% file: Tables/table_msa.tex
\begin{table}[!t]
\scriptsize
\setlength{\tabcolsep}{2pt} % default is 6pt

\centering
\caption{Comparative analysis -- MSA.}
\label{tab:msa}
\renewcommand\arraystretch{1}
\begin{tabular}{|c|ccc|c|c|cc|cc|c|}
\hline
 &
\multicolumn{3}{c|}{\textbf{Benign}} &
 & &
\multicolumn{5}{c|}{\textbf{Adversarial}}\\
\cline{2-4}\cline{7-11}
\textbf{Model} &
\textbf{FP} & \textbf{FN} & \textbf{MCC} &
\textbf{Attack} & \boldmath$\epsilon$ &
\textbf{FP} & {\bf ASR} & \textbf{FN} & \textbf{ASR} & \textbf{MCC}\\

\hline
\multirow{6}{*}{DT} & 1031 & 570 & 0.962 & FGSM & 1 & 1002 & 0.003 & 13 & 0.001 & 0.86\\
 & 1031 & 570 & 0.962 & FGSM & 5 & 1002 & 0.003 & 13 & 0.001 & 0.838\\
 & 1031 & 570 & 0.962 & BIM & 1 & 1002 & 0.003 & 13 & 0.001 & 0.872\\
 & 1031 & 570 & 0.962 & BIM & 5 & 1002 & 0.003 & 13 & 0.001 & 0.879\\
 & 1031 & 570 & 0.962 & PGD & 1 & 1124 & 0.003 & 3602 & 0.243 & 0.773\\
 & 1031 & 570 & 0.962 & PGD & 5 & 2830 & 0.008 & 6690 & 0.451 & 0.638\\
\hline
\multirow{6}{*}{RF} & 989 & 486 & 0.96 & FGSM & 1 & 961 & 0.003 & 24 & 0.002 & 0.858\\
 & 989 & 486 & 0.96 & FGSM & 5 & 961 & 0.003 & 24 & 0.002 & 0.855\\
 & 989 & 486 & 0.96 & BIM & 1 & 961 & 0.003 & 24 & 0.002 & 0.851\\
 & 989 & 486 & 0.96 & BIM & 5 & 961 & 0.003 & 24 & 0.002 & 0.858\\
 & 989 & 486 & 0.96 & PGD & 1 & 632 & 0.002 & 4418 & 0.298 & 0.763\\
 & 989 & 486 & 0.96 & PGD & 5 & 751 & 0.002 & 7835 & 0.528 & 0.66\\
\hline
\multirow{6}{*}{ET} & 1026 & 584 & 0.961 & FGSM & 1 & 998 & 0.003 & 276 & 0.019 & 0.855\\
 & 1026 & 584 & 0.961 & FGSM & 5 & 998 & 0.003 & 112 & 0.008 & 0.854\\
 & 1026 & 584 & 0.961 & BIM & 1 & 998 & 0.003 & 276 & 0.019 & 0.859\\
 & 1026 & 584 & 0.961 & BIM & 5 & 998 & 0.003 & 112 & 0.008 & 0.867\\
 & 1026 & 584 & 0.961 & PGD & 1 & 165 & 0.0 & 2217 & 0.149 & 0.885\\
 & 1026 & 584 & 0.961 & PGD & 5 & 283 & 0.001 & 2851 & 0.192 & 0.887\\
\hline
\multirow{6}{*}{XGB} & 1007 & 442 & 0.956 & FGSM & 1 & 980 & 0.003 & 13 & 0.001 & 0.858\\
 & 1007 & 442 & 0.956 & FGSM & 5 & 980 & 0.003 & 13 & 0.001 & 0.833\\
 & 1007 & 442 & 0.956 & BIM & 1 & 980 & 0.003 & 13 & 0.001 & 0.863\\
 & 1007 & 442 & 0.956 & BIM & 5 & 980 & 0.003 & 13 & 0.001 & 0.873\\
 & 1007 & 442 & 0.956 & PGD & 1 & 249 & 0.001 & 7360 & 0.496 & 0.648\\
 & 1007 & 442 & 0.956 & PGD & 5 & 1754 & 0.005 & 7468 & 0.503 & 0.63\\
\hline
\multirow{6}{*}{DNN} & 237 & 4226 & 0.855 & FGSM & 1 & 183 & 0.001 & 3152 & 0.212 & 0.882\\
 & 237 & 4226 & 0.855 & FGSM & 5 & 183 & 0.001 & 3152 & 0.212 & 0.783\\
 & 237 & 4226 & 0.855 & BIM & 1 & 183 & 0.001 & 3152 & 0.212 & 0.882\\
 & 237 & 4226 & 0.855 & BIM & 5 & 183 & 0.001 & 3152 & 0.212 & 0.882\\
 & 237 & 4226 & 0.855 & PGD & 1 & 182 & 0.001 & 3152 & 0.212 & 0.882\\
 & 237 & 4226 & 0.855 & PGD & 5 & 1475 & 0.004 & 3156 & 0.213 & 0.863\\
\hline
\end{tabular}
\end{table}

%% file: Tables/table_rloffa.tex
\begin{table}[!t]
\scriptsize
\setlength{\tabcolsep}{2.5pt} % default is 6pt
\centering
\caption{Comparative analysis -- RLOFFA.}
\label{tab:rloffa}
\renewcommand\arraystretch{1}
\begin{tabular}{|c|ccc|c|c|cc|cc|c|}
\hline
 &
\multicolumn{3}{c|}{\textbf{Benign}} &
 & &
\multicolumn{5}{c|}{\textbf{Adversarial}}\\
\cline{2-4}\cline{7-11}
\textbf{Model} &
\textbf{FP} & \textbf{FN} & \textbf{MCC} &
\textbf{Attack} & \boldmath$\epsilon$ &
\textbf{FP} & {\bf ASR} & \textbf{FN} & \textbf{ASR} & \textbf{MCC}\\
\hline

\multirow{6}{*}{DT} & 18 & 4590 & 0.733 & FGSM & 1 & 6 & 0.0 & 2799 & 0.438 & 0.34\\
 & 18 & 4590 & 0.733 & FGSM & 5 & 6 & 0.0 & 2799 & 0.438 & 0.401\\
 & 18 & 4590 & 0.733 & BIM & 1 & 6 & 0.0 & 2799 & 0.438 & 0.349\\
 & 18 & 4590 & 0.733 & BIM & 5 & 6 & 0.0 & 2799 & 0.438 & 0.35\\
 & 18 & 4590 & 0.733 & PGD & 1 & 535 & 0.004 & 5175 & 0.81 & 0.188\\
 & 18 & 4590 & 0.733 & PGD & 5 & 492 & 0.004 & 6096 & 0.954 & 0.129\\
\hline
\multirow{6}{*}{RF} & 18 & 4650 & 0.738 & FGSM & 1 & 6 & 0.0 & 2859 & 0.448 & 0.443\\
 & 18 & 4650 & 0.738 & FGSM & 5 & 6 & 0.0 & 2859 & 0.448 & 0.439\\
 & 18 & 4650 & 0.738 & BIM & 1 & 6 & 0.0 & 2859 & 0.448 & 0.454\\
 & 18 & 4650 & 0.738 & BIM & 5 & 6 & 0.0 & 2859 & 0.448 & 0.454\\
 & 18 & 4650 & 0.738 & PGD & 1 & 25 & 0.0 & 5373 & 0.841 & 0.231\\
 & 18 & 4650 & 0.738 & PGD & 5 & 54 & 0.0 & 6368 & 0.997 & 0.029\\
\hline
\multirow{6}{*}{ET} & 18 & 4590 & 0.733 & FGSM & 1 & 6 & 0.0 & 2799 & 0.438 & 0.423\\
 & 18 & 4590 & 0.733 & FGSM & 5 & 6 & 0.0 & 2799 & 0.438 & 0.413\\
 & 18 & 4590 & 0.733 & BIM & 1 & 6 & 0.0 & 2799 & 0.438 & 0.423\\
 & 18 & 4590 & 0.733 & BIM & 5 & 6 & 0.0 & 2799 & 0.438 & 0.423\\
 & 18 & 4590 & 0.733 & PGD & 1 & 24 & 0.0 & 5354 & 0.838 & 0.166\\
 & 18 & 4590 & 0.733 & PGD & 5 & 64 & 0.001 & 6365 & 0.997 & 0.025\\
\hline
\multirow{6}{*}{XGB} & 18 & 4593 & 0.742 & FGSM & 1 & 6 & 0.0 & 2799 & 0.438 & 0.45\\
 & 18 & 4593 & 0.742 & FGSM & 5 & 6 & 0.0 & 2799 & 0.438 & 0.439\\
 & 18 & 4593 & 0.742 & BIM & 1 & 6 & 0.0 & 2799 & 0.438 & 0.605\\
 & 18 & 4593 & 0.742 & BIM & 5 & 6 & 0.0 & 2799 & 0.438 & 0.602\\
 & 18 & 4593 & 0.742 & PGD & 1 & 27 & 0.0 & 5642 & 0.883 & 0.148\\
 & 18 & 4593 & 0.742 & PGD & 5 & 135 & 0.001 & 6333 & 0.992 & 0.031\\
\hline
\multirow{6}{*}{DNN} & 6 & 6687 & 0.612 & FGSM & 1 & 6 & 0.0 & 4566 & 0.715 & 0.371\\
 & 6 & 6687 & 0.612 & FGSM & 5 & 0 & 0.0 & 4566 & 0.715 & 0.153\\
 & 6 & 6687 & 0.612 & BIM & 1 & 6 & 0.0 & 4566 & 0.715 & 0.372\\
 & 6 & 6687 & 0.612 & BIM & 5 & 0 & 0.0 & 4566 & 0.715 & 0.372\\
 & 6 & 6687 & 0.612 & PGD & 1 & 3 & 0.0 & 4624 & 0.724 & 0.394\\
 & 6 & 6687 & 0.612 & PGD & 5 & 72 & 0.001 & 5465 & 0.856 & 0.318\\
\hline
\end{tabular}
\end{table}

%% file: Tables/table_rlona.tex
\begin{table}[!t]
\scriptsize
\setlength{\tabcolsep}{2pt} % default is 6pt

\centering
\caption{Comparative analysis -- RLONA.}
\label{tab:rlona}
\renewcommand\arraystretch{1}
\begin{tabular}{|c|ccc|c|c|cc|cc|c|}
\hline
 &
\multicolumn{3}{c|}{\textbf{Benign}} &
 & &
\multicolumn{5}{c|}{\textbf{Adversarial}}\\
\cline{2-4}\cline{7-11}
\textbf{Model} &
\textbf{FP} & \textbf{FN} & \textbf{MCC} &
\textbf{Attack} & \boldmath$\epsilon$ &
\textbf{FP} & {\bf ASR} & \textbf{FN} & \textbf{ASR} & \textbf{MCC}\\
\hline

\multirow{6}{*}{DT} & 1737 & 273 & 0.93 & FGSM & 1 & 2025 & 0.006 & 273 & 0.018 & 0.117\\
 & 1737 & 273 & 0.93 & FGSM & 5 & 2025 & 0.006 & 273 & 0.018 & 0.099\\
 & 1737 & 273 & 0.93 & BIM & 1 & 2025 & 0.006 & 273 & 0.018 & 0.133\\
 & 1737 & 273 & 0.93 & BIM & 5 & 2025 & 0.006 & 273 & 0.018 & 0.131\\
 & 1737 & 273 & 0.93 & PGD & 1 & 3327 & 0.01 & 8936 & 0.588 & 0.047\\
 & 1737 & 273 & 0.93 & PGD & 5 & 4497 & 0.014 & 14433 & 0.95 & 0.022\\
\hline
\multirow{6}{*}{RF} & 1824 & 258 & 0.93 & FGSM & 1 & 1896 & 0.006 & 258 & 0.017 & 0.206\\
 & 1824 & 258 & 0.93 & FGSM & 5 & 1896 & 0.006 & 258 & 0.017 & 0.161\\
 & 1824 & 258 & 0.93 & BIM & 1 & 1896 & 0.006 & 258 & 0.017 & 0.221\\
 & 1824 & 258 & 0.93 & BIM & 5 & 1896 & 0.006 & 258 & 0.017 & 0.205\\
 & 1824 & 258 & 0.93 & PGD & 1 & 2289 & 0.007 & 5961 & 0.393 & 0.144\\
 & 1824 & 258 & 0.93 & PGD & 5 & 1665 & 0.005 & 14075 & 0.927 & 0.037\\
\hline
\multirow{6}{*}{ET} & 1737 & 273 & 0.93 & FGSM & 1 & 1653 & 0.005 & 273 & 0.018 & 0.851\\
 & 1737 & 273 & 0.93 & FGSM & 5 & 1809 & 0.005 & 273 & 0.018 & 0.263\\
 & 1737 & 273 & 0.93 & BIM & 1 & 1653 & 0.005 & 273 & 0.018 & 0.858\\
 & 1737 & 273 & 0.93 & BIM & 5 & 1809 & 0.005 & 273 & 0.018 & 0.843\\
 & 1737 & 273 & 0.93 & PGD & 1 & 1185 & 0.004 & 2210 & 0.146 & 0.831\\
 & 1737 & 273 & 0.93 & PGD & 5 & 918 & 0.003 & 7644 & 0.503 & 0.487\\
\hline
\multirow{6}{*}{XGB} & 1899 & 324 & 0.93 & FGSM & 1 & 2025 & 0.006 & 324 & 0.021 & 0.112\\
 & 1899 & 324 & 0.93 & FGSM & 5 & 2025 & 0.006 & 324 & 0.021 & 0.093\\
 & 1899 & 324 & 0.93 & BIM & 1 & 2025 & 0.006 & 324 & 0.021 & 0.139\\
 & 1899 & 324 & 0.93 & BIM & 5 & 2025 & 0.006 & 324 & 0.021 & 0.136\\
 & 1899 & 324 & 0.93 & PGD & 1 & 2005 & 0.006 & 8127 & 0.535 & 0.038\\
 & 1899 & 324 & 0.93 & PGD & 5 & 3173 & 0.01 & 14186 & 0.934 & 0.054\\
\hline
\multirow{6}{*}{DNN} & 702 & 1674 & 0.797 & FGSM & 1 & 462 & 0.001 & 1665 & 0.11 & 0.819\\
 & 702 & 1674 & 0.797 & FGSM & 5 & 462 & 0.001 & 1665 & 0.11 & 0.12\\
 & 702 & 1674 & 0.797 & BIM & 1 & 462 & 0.001 & 1665 & 0.11 & 0.826\\
 & 702 & 1674 & 0.797 & BIM & 5 & 462 & 0.001 & 1665 & 0.11 & 0.827\\
 & 702 & 1674 & 0.797 & PGD & 1 & 1013 & 0.003 & 3140 & 0.207 & 0.837\\
 & 702 & 1674 & 0.797 & PGD & 5 & 1217 & 0.004 & 6916 & 0.455 & 0.578\\
\hline
\end{tabular}
\end{table}

%% file: Tables/table_csa.tex
\begin{table}[!t]
\setlength{\tabcolsep}{2.5pt} % default is 6pt
\scriptsize
\centering
\caption{Comparative analysis -- CSA.}
\label{tab:csa}
\renewcommand\arraystretch{1}
\begin{tabular}{|c|c|c|c|c|ccc|}
\hline
\textbf{Model} & \multicolumn{2}{c|}{\textbf{Benign}} & \textbf{Attack} & \boldmath\textbf{$\epsilon$} & \multicolumn{3}{c|}{\textbf{Adversarial}}\\
\cline{2-3}\cline{6-8}
 & \textbf{F1 Score} & \textbf{FP} &  &  & \textbf{F1 Score} & \textbf{FP} & \textbf{ASR}\\
\hline
\multirow{6}{*}{DT} & 1.000 & 99 & FGSM & 1 & 0.999 & 137 & 0.001\\
 & 1.000 & 99 & FGSM & 5 & 0.999 & 137 & 0.001\\
 & 1.000 & 99 & BIM & 1 & 0.999 & 137 & 0.001\\
 & 1.000 & 99 & BIM & 5 & 0.999 & 137 & 0.001\\
 & 1.000 & 99 & PGD & 1 & 0.999 & 214 & 0.002\\
 & 1.000 & 99 & PGD & 5 & 0.998 & 440 & 0.004\\
\hline
\multirow{6}{*}{RF} & 1.000 & 97 & FGSM & 1 & 1.000 & 48 & 0.000\\
 & 1.000 & 97 & FGSM & 5 & 1.000 & 48 & 0.000\\
 & 1.000 & 97 & BIM & 1 & 1.000 & 48 & 0.000\\
 & 1.000 & 97 & BIM & 5 & 1.000 & 48 & 0.000\\
 & 1.000 & 97 & PGD & 1 & 1.000 & 27 & 0.000\\
 & 1.000 & 97 & PGD & 5 & 1.000 & 65 & 0.001\\
\hline
\multirow{6}{*}{ET} & 1.000 & 97 & FGSM & 1 & 1.000 & 47 & 0.000\\
 & 1.000 & 97 & FGSM & 5 & 1.000 & 47 & 0.000\\
 & 1.000 & 97 & BIM & 1 & 1.000 & 47 & 0.000\\
 & 1.000 & 97 & BIM & 5 & 1.000 & 47 & 0.000\\
 & 1.000 & 97 & PGD & 1 & 1.000 & 9 & 0.000\\
 & 1.000 & 97 & PGD & 5 & 1.000 & 19 & 0.000\\
\hline
\multirow{6}{*}{XGB} & 1.000 & 115 & FGSM & 1 & 0.999 & 137 & 0.001\\
 & 1.000 & 115 & FGSM & 5 & 0.999 & 137 & 0.001\\
 & 1.000 & 115 & BIM & 1 & 0.999 & 137 & 0.001\\
 & 1.000 & 115 & BIM & 5 & 0.999 & 137 & 0.001\\
 & 1.000 & 115 & PGD & 1 & 1.000 & 51 & 0.000\\
 & 1.000 & 115 & PGD & 5 & 0.999 & 191 & 0.002\\
\hline
\multirow{6}{*}{DNN} & 0.999 & 322 & FGSM & 1 & 1.000 & 69 & 0.001\\
 & 0.999 & 322 & FGSM & 5 & 1.000 & 69 & 0.001\\
 & 0.999 & 322 & BIM & 1 & 1.000 & 69 & 0.001\\
 & 0.999 & 322 & BIM & 5 & 1.000 & 69 & 0.001\\
 & 0.999 & 322 & PGD & 1 & 1.000 & 59 & 0.000\\
 & 0.999 & 322 & PGD & 5 & 1.000 & 93 & 0.001\\
\hline
\end{tabular}

\begin{tabular}{|c|}
\hline
\textit{No FN results for CSA were produced in \texttt{csa\_fn.csv}.} \\
\hline
\end{tabular}
\end{table}

%% file: 6_conclusion.tex
\section{Conclusion}

This work presents a systematic adversarial evaluation of in-vehicle IDS models operating on the CAN bus using the ROAD dataset. By analyzing both shallow and DL classifiers, we assess how different model architectures respond to constrained and realistic adversarial perturbations. The results reveal clear differences in robustness: Despite the DNN performing best on benign traffic, gradient-based attacks including FGSM, BIM, and PGD fully compromise the DNN-based IDS in terms of missed attacks under specific scenarios, whereas ensemble models, particularly ET, exhibit comparatively stronger resilience. These findings indicate that although neural models can capture complex feature relationships, they are highly susceptible to gradient-driven adversarial manipulation. Moreover, enforcing strict CAN domain constraints by limiting perturbations to valid payload bytes demonstrates that adversarial attacks can remain protocol-compliant and stealthy, underscoring their practical feasibility.

Overall, our results highlight the importance of adversarial robustness evaluation in automotive cybersecurity research. Future work will explore defense strategies such as adversarial training, input sanitization, and hybrid IDS designs that combine learned representations with interpretable structures. As vehicles continue to advance toward greater autonomy and connectivity, ensuring that IDS solutions remain robust against adaptive adversaries will be critical for securing next-generation intelligent transportation systems.

%% file: ack.tex
\section*{Acknowledgment}
 This manuscript has been authored by UT-Battelle, LLC under Contract No. DE-AC05-00OR22725 with the U.S. Department of Energy. The publisher, by accepting the article for publication, acknowledges that the U.S. Government retains a non-exclusive, paid-up, irrevocable, worldwide license to publish or reproduce the published form of the manuscript, or allow others to do so, for U.S. Government purposes. The DOE will provide public access to these results in accordance with the DOE Public Access Plan. There was no additional external funding received for this study. The funders had no role in study design, data collection and analysis, decision to publish, or preparation of this manuscript. (\url{http://energy.gov/downloads/doe-public-access-plan}). This research was sponsored in part by Oak Ridge National Laboratory’s (ORNL’s) Laboratory Directed Research and Development program.